# Upconversion time-stretch infrared spectroscopy

Kazuki Hashimoto[1], Takuma Nakamura[1], Takahiro Kageyama[2], Venkata Ramaiah Badarla[1], Hiroyuki Shimada[1], Ryoich Horisaki[3], and Takuro Ideguchi[1,2,*]

[1] Institute for Photon Science and Technology, The University of Tokyo, Tokyo, Japan
[2] Department of Physics, The University of Tokyo, Tokyo, Japan
[3] Graduate School of Information Science and Technology, The University of Tokyo, Tokyo, Japan
*ideguchi@ipst.s.u-tokyo.ac.jp

**High-speed measurement confronts the extreme speed limit when the signal becomes comparable to the noise level. In the context of broadband mid-infrared spectroscopy[1], state-of-the-art ultrafast Fourier-transform infrared spectrometers[2–14], in particular dual-comb spectrometers, have improved the measurement rate up to a few Mspectra/s[6,8,12], which is limited by the signal-to-noise ratio. Time-stretch infrared spectroscopy, an emerging ultrafast frequency-swept mid-infrared spectroscopy technique, has shown a record-high rate of 80 Mspectra/s with an intrinsically higher signal-to-noise ratio than Fourier-transform spectroscopy by more than the square-root of the number of spectral elements[15]. However, it can measure no more than ~30 spectral elements with a low resolution of several $cm^{-1}$. Here, we significantly increase the measurable number of spectral elements to more than 1,000 by incorporating a nonlinear upconversion process[16–19]. The one-to-one mapping of a broadband spectrum from the mid-infrared to the near-infrared telecommunication region enables low-loss time-stretching with a single-mode optical fiber and low-noise signal detection with a high-bandwidth photoreceiver. We demonstrate high-resolution mid-infrared spectroscopy of gas-phase methane molecules with a high resolution of 0.017 $cm^{-1}$. This unprecedentedly high-speed vibrational spectroscopy technique would satisfy various unmet needs in experimental molecular science, e.g., measuring ultrafast dynamics of irreversible phenomena[12,13,20], statistically analyzing a large amount of heterogeneous spectral data[21,22], or taking broadband hyperspectral images at a high frame rate[23–25].**

Broadband mid-infrared (MIR) spectroscopy is a powerful non-invasive tool for identifying molecular species and sensing subtle changes in molecular structures that reflect environmental conditions, and various applications have been widely investigated, such as environmental gas monitoring[14,26,27], combustion analysis[20,28], photoreactive protein analysis[12], liquid biopsy[21,29], breath diagnosis[30,31], etc. One of the promising directions of instrumental development of broadband MIR spectroscopy is increasing the measurement speed because normal vibration modes of molecules have huge MIR absorption cross-sections, which are orders of magnitude larger than Raman scattering cross-sections. One approach for that is a parallel signal detection using a sensor array with a grating-based dispersive spectrometer, but the spectral measurement rate is limited by the readout rate of the sensor, which is typically up to ~1 kspectra/s[32,33]. Another approach is increasing the scan rate of a Fourier-transform spectrometer by taking advantage of the high bandwidth of a photodetector. Advanced FTIR spectrometers such as MIR dual-comb spectroscopy (MIR-DCS)[2–9,12–14], rapid-scan FTIR[10], and phase-controlled FTIR[11] have remarkably improved the measurement scan rate up to a few Mspectra/s[6,8,12]. These high-speed FTIR techniques could open doors for applications such as measuring non-repetitive rapid phenomena at high temporal resolution and making statistical analyses of a significantly large amount of spectral data. However, the measurement rate has already hit the theoretical maximum limited by the signal-to-noise ratio (SNR). Therefore, to improve the measurement rate further, one needs a method with a fundamentally higher SNR.

Frequency-swept spectroscopy (FSS)[34–36] is a method having higher SNR than FTIR, where a broadband spectrum is measured by sweeping the laser frequency. It has more than $\sim\sqrt{M}$ ($M$: number of spectral elements) times higher SNR than FTIR[37] due to the less noise per spectral element (theoretical description is shown in Supplementary Note 8). Therefore, FSS techniques have the potential to increase the spectral measurement rate, but, to the best of our knowledge, the highest scan rate of MIR-FSS with a frequency-swept MIR laser was 250 kspectra/s[35], which can measure two spectral elements only. We proposed that a time-stretched pulse could be used as a high-speed frequency-swept laser and demonstrated time-stretch infrared spectroscopy (TSIR) at the record spectral measurement rate of 80 Mspectra/s[15]. Although the developed system significantly improved the measurement rate, the measurable number of spectral elements was limited to about 30 (the spectral resolution was 7.7 cm$^{-1}$), mainly due to the large loss in time-stretching with a free-space angular-chirp-enhanced delay line (FACED)[38] and the low sensitivity in MIR photodetection. With the small number of spectral elements, one does not gain the full advantage of broadband MIR spectroscopy, i.e., the high SNR multiplex spectral measurement with a large number of spectral elements. Furthermore, the low spectral resolution does not allow one to apply it to gas-phase spectroscopy.

In this work, we develop upconversion TSIR (UC-TSIR) and demonstrate high-speed and high-resolution broadband MIR spectroscopy with spectral elements more than 1,000 at a rate of above 10 Mspectra/s. The nonlinear upconversion[16–19] via difference frequency generation (DFG) allows the implementation of time-stretching and photodetection in the near-infrared telecommunication region, where high-quality optics and optical devices are well developed. It provides superior advantages for TSIR: (1) low-loss and large pulse-stretching with a telecommunication-grade optical fiber[39,40] and (2) low-noise and high-bandwidth pulse detection with an InGaAs photodetector, enabling high-speed, high-resolution, and high-content MIR spectroscopy. As a proof of concept demonstration, we measure gaseous $CH_4$ molecules with different pulse-stretching conditions. For a demonstration of high-speed capability, it is operated at 80 Mspectra/s with a spectral resolution of 0.10 cm$^{-1}$ and -10-dB bandwidth of 20 cm$^{-1}$ (200 spectral elements). The threshold level of spectral bandwidth (-10 dB in this case) is determined where the single-shot SNR becomes 1. For a demonstration of high spectral resolution, it is operated at a rate of 10 Mspectra/s with spectral resolutions and bandwidths of 0.034 cm$^{-1}$ and 26 cm$^{-1}$ at -10 dB level (760 spectral elements), and 0.017 cm$^{-1}$ and 17 cm$^{-1}$ at -8 dB level (1,000 spectral elements).

Figure 1 illustrates a comparison of the working principle and SNR between TSIR and FTIR. A TSIR spectrometer measures spectra directly in the time domain by photonic time-stretch, also known as dispersive Fourier-transformation (DFT), while an FTIR spectrometer measures temporal interferograms and converts them to spectra by fast Fourier-transformation (FFT). In TSIR, the noise of a sampled data point directly becomes that of a corresponding spectral element, whereas, in FTIR, the noise of all the data points of an interferogram contributes to that of a spectral element via FFT. The difference in the amount of noise per spectral element results in the $\sqrt{M}$-times higher SNR of TSIR, where $M$ is the number of spectral elements. In addition, TSIR allows measuring a two-averaged spectrum within a measurement time of a single FTIR spectrum because the number of FTIR spectral elements is half the number of the sampling data points due to the Nyquist theorem, which gives an additional SNR factor of $\sqrt{2}$. Furthermore, under the condition where the detector's dynamic range limits the SNR, TSIR obtains an extra SNR advantage of a factor $\alpha$ ($1 \leq \alpha \leq 2$), which is a constant value determined by noise conditions. It comes from the FTIR's working mechanism, where the DC signal of the interferogram consumes half of the dynamic range. In total, TSIR has $\alpha\sqrt{2M}$-times higher SNR over FTIR. The detailed theoretical description is summarized in Supplementary Note 8.

Figures 1b and c show the theoretically calculated SNR of TSIR (Equation S12) and FTIR (Equation S21) at $\alpha = 1$ (the lowest value of $\alpha$) as a function of the number of spectral elements $M$ and measurement time $T$, respectively. Figure 1b visualizes TSIR has $\sqrt{2M}$-times higher SNR than FTIR, and the SNR advantage becomes significant when $M$ is large. For example, TSIR can have higher SNR than FTIR by 45 when $M=1,000$. Note that the vertical axis of the SNR is an arbitrary unit, which varies depending on measurement conditions, including measurement time. Figure 1c shows the SNR dependence on measurement time $T$ for $M=1,000$. Here, we calculate the SNR by assuming flat-top spectra under a measurement condition with typical parameter values: the average detection power of 10 µW, the system's overall noise-equivalent power of 10 pW/$\sqrt{Hz}$. We assume that the detection bandwidth and the sampling rate are sufficiently high for the measurement. The SNR of TSIR becomes 1 at the measurement time of 500 ps, showing the potential to achieve a measurement rate of 2 Gspectra/s. On the other hand, FTIR requires 1 µs to achieve an SNR of 1, limiting the maximum measurement rate at 1 Mspectra/s, which agrees well with the results of the previous works of high-speed FTIR such as MIR-DCS [6,8,12].

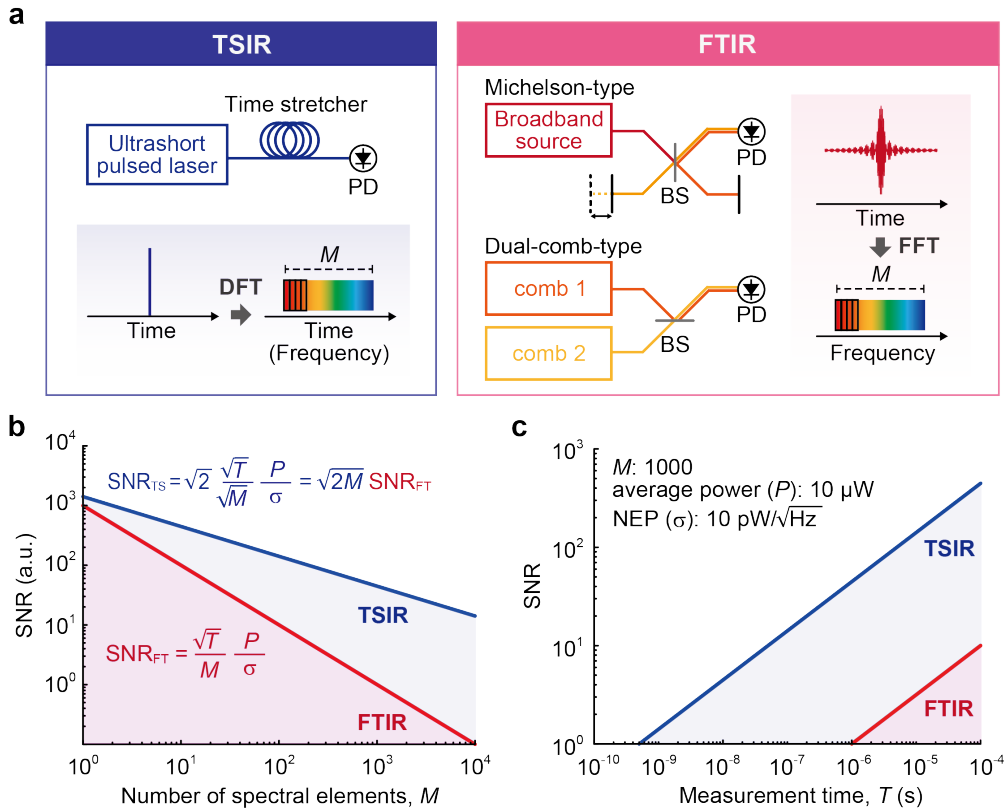

**Figure 1:** Comparison of working principle and SNR between TSIR and FTIR. **a,** Working principle of TSIR and FTIR. DFT: dispersive Fourier-transform, FFT: Fast-Fourier transform. **b,** SNR dependence on the number of spectral elements, $M$ (Blue: TSIR, Red: FTIR). $P$: average detection power, $\sigma$: system's overall noise equivalent power. **c,** SNR dependence on measurement time, $T$ (Blue: TSIR, Red: FTIR). The SNR is calculated for $M=1,000$ with typical values of $P= 10$ µW and $\sigma=10$ pW/$\sqrt{Hz}$.

Figure 2 illustrates a schematic of our UC-TSIR spectrometer. We use a homemade 80-MHz femtosecond MIR optical parametric oscillator (OPO) with a -10-dB spectral bandwidth of 235 cm$^{-1}$ at a center wavelength of 3.47 µm as a broadband

MIR light source. The MIR beam passing through a sample (a $CH_4$ gas cell) with an average power of a few mW is spatially combined with a 1.064-μm continuous-wave (CW) laser beam with an average power of around 400 mW. The combined beams are focused onto a 20-mm-long periodically-poled lithium niobate (PPLN) waveguide with an aspheric lens. The DFG process in the waveguide converts the MIR pulses to NIR pulses around 1.5 μm with an average power of 3 μW. The -10-dB spectral bandwidth of the upconverted NIR pulses is 21 $cm^{-1}$ (5.1 nm), which is determined by the phase-matching condition in the PPLN waveguide. The use of the 1-μm CW laser for the DFG guarantees a one-to-one spectral transfer from a MIR pulse to a NIR pulse. The generated NIR pulses are coupled into a single-mode fiber with a coupling efficiency of 0.65 and optically amplified with an Er-doped fiber amplifier (EDFA). Subsequently, they are temporally stretched by dispersion-compensating fibers (DCF) with a total length of 10, 30, or 60 km. The DCF's dispersion parameter is -0.2 ns/nm/km. In large-stretching cases with a fiber length of 30 and 60 km (total dispersion of -6 and -12 ns/nm, respectively), we additionally implement a pulse picker to avoid temporal overlap of adjacent pulses and a Raman amplifier to retain the signal intensity. The stretched NIR pulses pass through optical bandpass filters for spectral bandwidth adjustment. The filtered NIR pulses are detected and digitized with an 11-GHz InGaAs photodetector and a 16-GHz oscilloscope at a sampling rate of 80 Gsamples/s. The details of the system are described in Methods and Supplementary Note 1. To avoid spectral distortions due to undesired nonlinear effects, it is essential to carefully manage the pulse energy in the PPLN waveguide and the optical fiber. The details of the nonlinear effects are described in Supplementary Notes 2 and 3.

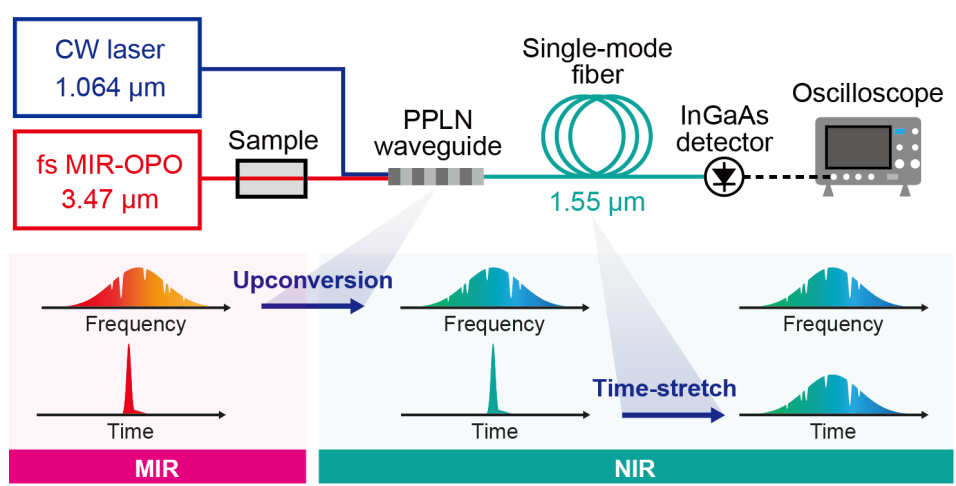

**Figure 2:** Schematic of upconversion TSIR (UC-TSIR). MIR-OPO: mid-infrared optical parametric oscillator, PPLN: periodically-poled lithium niobate.

We demonstrate broadband UC-TSIR spectroscopy of gaseous $CH_4$ molecules in a 50-mm-long cell with a pressure of 10 Torr. To show the high-speed capability, we first operate the TSIR spectrometer at 80 Mspactra/s by setting the fiber length to 10 km (dispersion of -2 ns/nm) to stretch the spectrum over the pulses' interval of 12.5 ns. The left panel in Fig. 3a shows temporal TSIR waveforms, that is, TSIR spectra. The TSIR spectral resolution is determined either by the pulse duration before stretching or the impulse response of the detector. In our experiments, the measured temporal width of an unstretched NIR pulse using the photodetector is 49 ps, which is determined by the impulse response of the detector (details are shown in Supplementary Note 4). Under this condition, the number of TSIR spectral elements is 200 with a -10-dB spectral bandwidth of 20 $cm^{-1}$ (corresponding to 9 ns in the time domain) and a spectral resolution of 0.10 $cm^{-1}$ (49 ps). Absorption lines of gaseous $CH_4$ molecules appear on the spectrum, verifying the capability of high-speed and high-contents broadband MIR spectroscopy. The distortions in the absorption lineshape are not due to the measurement noise but systematic interference resulting from the near-field effect of dispersive propagation in the long fiber. The near-field propagation effect can be analogically explained as a temporal version of the well-known spatial near-field diffraction pattern, known as the Fresnel diffraction pattern (See Supplementary Note 5 for details).

Next, we use longer optical fibers of 30 km (dispersion of -6 ns/nm) and 60 km (dispersion of -12 ns/nm) to demonstrate higher spectral resolution (the middle and right panels in Fig. 3a, respectively). The pulse repetition rate (TSIR spectral measurement rate) is set to 10 MHz with a pulse picker to avoid temporal overlaps of the stretched adjacent pulses. To keep a sufficient signal intensity throughout the long travel in the optical fiber, we implement a Raman amplifier in the stretching fiber to compensate for the fiber loss, e.g., 22 dB in a 30-km DCF. Under this condition, the number of spectral elements for 30-km DCF is 760 with a -10-dB spectral bandwidth of 26 cm$^{-1}$ and a spectral resolution of 0.034 cm$^{-1}$, and that for 60-km DCF is 1,000 with a -8 dB spectral bandwidth of 17 cm$^{-1}$ and a spectral resolution of 0.017 cm$^{-1}$. We evaluate SNR by taking the standard deviation of a baseline-normalized single TSIR spectrum where large absorption lines do not exist. The baseline-normalized TSIR spectrum is calculated by dividing a measured TSIR spectrum by an envelope curve processed by Savitzky–Golay (SG) filtering[41]. The single-shot SNRs are 10 with the 30-km DCF and 6 with the 60-km DCF when the average detection powers are 21 μW and 12 μW, respectively. In our current system, the SNR is limited by the shot noise determined by the number of photons before the optical amplification and the amplified spontaneous emission (ASE) noise of the optical amplifiers for the average detection power above 10 μW. The detailed discussion about the SNR is described in Supplementary Note 9.

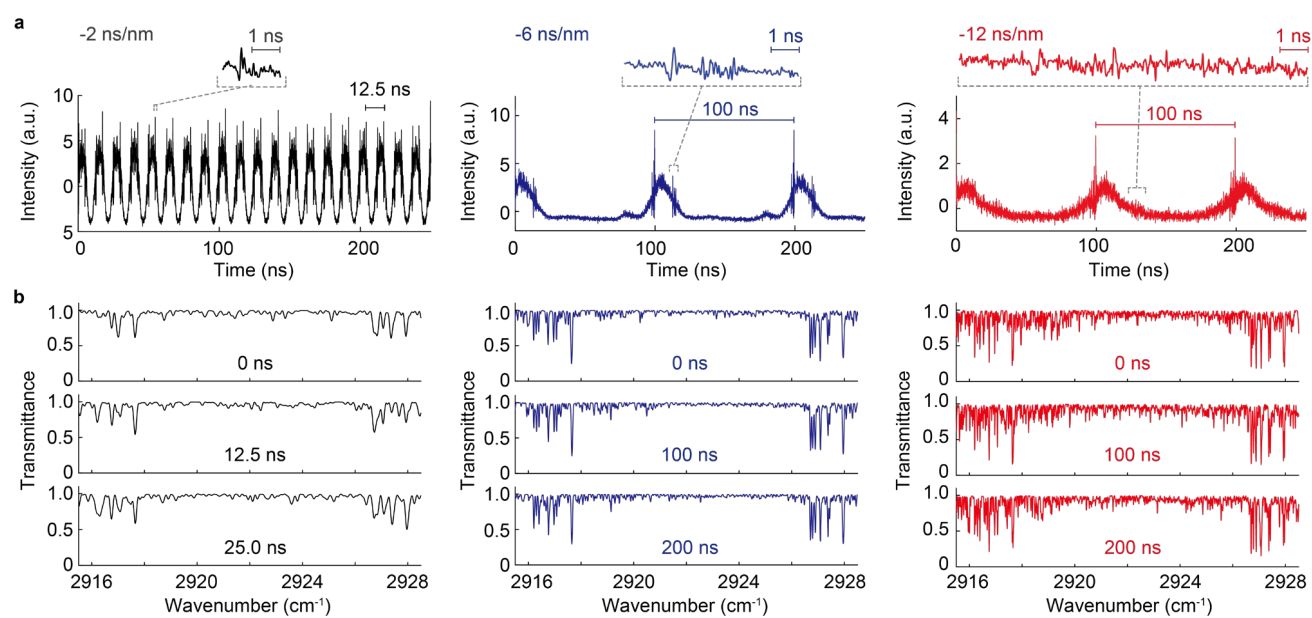

**Figure 3:** High-speed broadband MIR spectroscopy of gas-phase CH$_4$ molecules with upconversion TSIR. **a,** Continuously measured TSIR spectra at a rate of 80 Mspectra/s with a 10-km DCF (total dispersion of -2 ns/nm) (black), 10 Mspectra/s with a 30-km DCF (total dispersion of -6 ns/nm) (blue), and 10 Mspectra/s with a 60-km DCF (total dispersion of -12 ns/nm) (red). The insets show enlarged views of a part of the spectra. **b,** Single-shot transmittance spectra retrieved by the GD algorithm measured at 80 Mspectra/s with a 10-km DCF (black), 10 Mspectra/s with a 30-km DCF (blue), and 10 Mspectra/s with a 60-km DCF (red).

The systematic spectral distortions due to the near-field propagation can be demodulated by the iterative Gradient-descent (GD) algorithm[42]. The spectral retrieval procedures are described in Methods. Figure 3b shows demodulated transmittance spectra of $CH_4$ molecules retrieved from the TSIR spectra measured at 80 Mspectra/s with the 10-km DCF (total dispersion of -2 ns/nm), 10 Mspectra/s with the 30-km DCF (total dispersion of -6 ns/nm), and 10 M spectra/s with 60-km DCF (total dispersion of -12 ns/nm), respectively. The absorption lines of $CH_4$ molecules are recovered with a spectral resolution of 0.12 $cm^{-1}$ (3.6 GHz), 0.04 $cm^{-1}$ (1.2 GHz), and 0.02 $cm^{-1}$ (600 MHz), respectively. The wavenumber axis is downconverted from that in the NIR region, whose relative accuracy is determined by the dispersion values for pulse stretching used in the GD algorithm (see Methods and Supplementary Note 7). The values of group-delay dispersion (GDD) and third-order dispersion (TOD) used in the calculation are described in Methods. The GDD values are calibrated by comparing the measured and calculated TSIR spectra, while the TOD values are estimated from the relative dispersion slope of the DCF given in a product's datasheet.

Figure 4a shows 180-times averaged TSIR spectra with dispersions of -6 and -12 ns/nm. The pulses are stretched to 84 ns and 70 ns at -20-dB intensity level, which correspond to the spectral bandwidths of 59 $cm^{-1}$ (14 nm) and 24 $cm^{-1}$ (5.8 nm), respectively. We set the spectral bandwidth for each case with an optical bandpass filter. Considering the 49-ps impulse response of the photodetector, the number of TSIR spectral elements is 1,700 and 1,400, respectively. The averaging works well due to the high stability in the temporal axis. The standard deviation of the peak position is 9 ps, which is shorter than the oscilloscope's sampling time resolution of 12.5 ps. It is evaluated with a spectral point at 56.16 ns of the continuously measured single-shot TSIR spectra with a 30-km DCF (Details are discussed in Supplementary Note 6). Figure 4b compares the measured TSIR spectra and theoretically calculated spectra based on Equations (S1-S7) with parameters from the experiment and the HITRAN database[43]. The temporal baselines of the measured TSIR spectra are normalized by dividing the measured TSIR spectra by the envelope curve processed by SG filtering for a comparison with the calculated spectra. The spectral phase used in the calculation is deduced from the Kramers–Kronig (K-K) relation. The GDD and TOD values used in the calculation are 7,664 $ps^2$ and -48 $ps^3$ for the spectrum with the dispersion of -6 ns/nm, and 15,326 $ps^2$ and -96 $ps^3$ for the spectrum with the dispersion of -12 ns/nm, respectively. The figure shows that the measured spectra agree well with the calculated ones.

Figure 4c compares the 180-averaged measured transmittance spectrum (with a 60-km DCF) retrieved with the GD algorithm and the calculated transmittance spectrum from the HITRAN database (ground truth). The absorption lines of $CH_4$ molecules in the measured transmittance spectrum are in good agreement with the ground-truth spectrum at the spectral resolution of 0.02 $cm^{-1}$ (600 MHz). As seen in the inset of the figure, several a-few-% absorption peaks are clearly observed in the retrieved spectrum. There are relatively large residuals around, e.g., 2926-2928 $cm^{-1}$, which are likely caused by estimation error of spectral baseline because there are some bumps in the MIR baseline spectrum itself. We could suppress the residuals by using a flatter MIR spectrum without bumpy structures. Deviations of the peak positions of the absorption lines from the HITRAN are within 0.007 $cm^{-1}$ (200 MHz) (See Supplementary Note 7 for details).

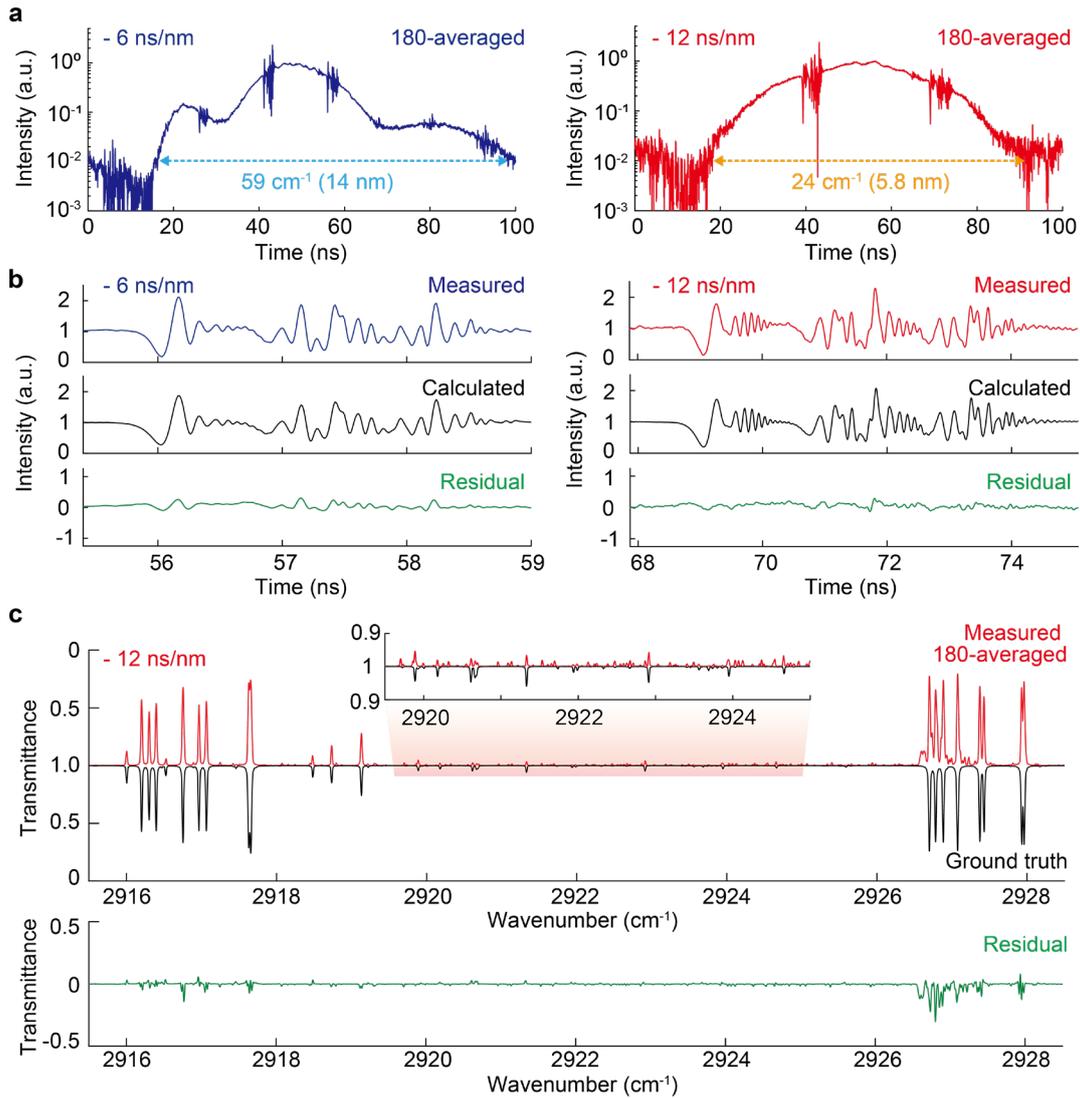

**Figure 4:** High-resolution broadband TSIR spectra of $CH_4$ molecules. **a,** 180-times averaged TSIR spectra measured with a 30-km DCF (dispersion of -6 ns/nm) (blue) and a 60-km DCF (dispersion of -12 ns/nm) (red). **b,** Comparison between a part of measured and calculated TSIR spectra. The temporal baselines of the measured TSIR spectra are normalized by the envelope functions. The green plots represent the residuals. **c,** A retrieved transmittance spectrum (dispersion of -12 ns/nm) by the GD algorithm (red) compared to a calculated ground-truth transmittance spectrum from the HITRAN database (black). The inset shows a zoom-in view of small absorption peaks. The green plot shows residual.

We make a detailed comparison between the UC-TSIR system and the previous TSIR system[15], particularly about the performance of their time-stretchers and photodetectors. The time-stretching in UC-TSIR is made with a low-loss DCF fiber in the NIR telecommunication region, while that in the previous TSIR is made with a FACED system in the MIR region. FACED has a large loss due to the multiple reflections on flat mirrors in free space. To make a comparison, for example, if we allow 10 dB loss for the stretching, they have abilities to add dispersions of -3 ns/nm (DCF) and -20 ps/nm (FACED), respectively. Therefore, the DCF time-stretcher can stretch a pulse by two-orders-of-magnitude longer. The fiber time-stretcher has another advantage in keeping spatial mode for a long-range (tens of km) due to the nature of waveguiding, while it is difficult to make the long-range propagation with FACED because of the beam divergence in free space. Furthermore, the fiber time-stretcher can work as an optical amplifier to compensate for the propagation loss, enabling the pulse-stretching even longer. Regarding photodetectors, UC-TSIR uses an amplified InGaAs photodetector with a responsivity of ~1 A/W, while the previous TSIR system uses a quantum cascade detector (QCD) working in the MIR region with a responsivity of ~10 mA/W[44]. The QCD's low responsivity requires more than a few mW photodetection for capturing the signals, resulting in additional difficulty in implementing a TSIR system. With these advantages, UC-TSIR enables broadband MIR spectroscopy with a significantly large number of spectral elements and a high-spectral resolution by maintaining high-speed capability.

For a comparison to the state-of-the-art high-speed MIR-FSS, we discuss the system performances of UC-TSIR and rapid-scan EC-QCL spectrometer[34–36,45] in terms of spectral measurement rate and the measurable number of spectral elements. The fastest EC-QCL spectrometer with an acousto-optic modulator (AOM) can be operated at a rate of 1 Mspectra/s for measuring two spectral elements only[36]. The scan rate is limited by the propagation speed of acoustic waves generated by a piezoelectric transducer. The AOM-based rapid-scan EC-QCL spectrometer demonstrates MIR spectroscopy with a bandwidth of 50 cm$^{-1}$ (two spectral elements) at a rate of 250 kHz[35] and 200 cm$^{-1}$ (~20 spectral elements) at 15 kHz[45]. On the other hand, UC-TSIR does not suffer from the speed limitation caused by active optical devices because of the ultrafast passive frequency sweep enabled by time-stretching. Therefore, it can measure the number of spectral elements more than $10^3$ at a measurement rate of tens of MHz, far exceeding the previous state-of-the-art.

The performance of UC-TSIR spectrometer can be improved further with system modifications. The SNR of the current system is limited by the shot noise determined by the number of photons before the optical amplification and the ASE noise from the optical amplifier. It can be improved by several times (Supplementary Note 9) by increasing the number of upconverted photons before the optical amplification, which improves the shot-noise-limited SNR determined by the number of photons before the optical amplification and can also reduce the optical amplification noise. Regarding measurable samples, UC-TSIR can be applied to broadband MIR spectroscopy of condensed media by expanding the spectral bandwidth. For example, the spectral bandwidth can be larger than hundreds of cm$^{-1}$ by using a few-mm-long PPLN crystal for upconversion. The foreseeing applications of the broadband UC-TSIR spectrometer are, for example, high-throughput single-cell analysis[22,46] or accurate molecular fingerprinting of bio-molecules for health monitoring[29]. The concept of wavelength-conversion TSIR can also be applied to other wavelength regions where low-loss time-stretchers do not exist. Although the current UC-TSIR system can be operated in a laboratory only due to the bulky fs MIR-OPO, it can be portable by using a compact and stable MIR source such as fiber-based lasers[16,18].

Finally, we discuss another spectroscopic aspect of UC-TSIR compared to FTIR, particularly the wavenumber calibration and the consequent accuracy. In UC-TSIR, it is necessary to calibrate the wavenumber scale by measuring the dispersion of optical fibers. We use a molecular absorption spectrum for the calibration, which limits the wavenumber accuracy. On the other hand, FTIR is capable of accurate interferometric calibration, e.g., with a HeNe laser in a Michelson-type FTIR. DCS can provide extremely high accuracy given by the nature of frequency combs with the level of atomic clocks. Therefore, FTS has an advantage in highly accurate precision spectroscopy, which is not within the main scope of the high-speed TSIR because precision measurement demands an extremely high SNR with a long measurement time.

In summary, we demonstrated UC-TSIR spectroscopy and showed high-speed broadband MIR spectroscopy of gas-phase molecules at an unprecedented level. By taking advantage of the superiority of the optical components and devices in the telecommunication region, we significantly improved the measurable number of spectral elements and the spectral resolution by maintaining the high SNR and measurement speed. The UC-TSIR spectrometer could enable various applications, particularly measurement of complex irreversible phenomena at a high temporal resolution[12,13,20] and statistical analysis of a large number of high contents spectral data[21,22,46]. It can also be applicable to other sensing techniques, such as MIR optical coherence tomography[47] for 3D deep-inside profiling of highly scattering media.

## Methods

### Light sources

We use a homemade fs MIR-OPO pumped by an 80-MHz Ti:Sapphire mode-locked laser (Maitai, Spectra-Physics) as a broadband MIR light source. The OPO generates MIR idler pulses with an average power of around 100 mW. In our experiment, the center wavenumber is adjusted to 2,880 $cm^{-1}$ (3.47 μm), whose -10-dB spectral bandwidth is about 235 $cm^{-1}$, as shown in Supplementary Figure 1a. The MIR pulses are coupled into an $InF_3$ single-mode fiber using an aspheric lens for spatial mode cleaning. A MIR bandpass filter with a bandwidth of 49 $cm^{-1}$ is installed before the fiber-coupling to suppress undesired nonlinear optical effects in the optical fiber and the PPLN waveguide for upconversion. The fiber-output MIR pulses are collimated with a collimator and pass through a sample. We use gaseous $CH_4$ molecules as a sample (CH4-T(25x5)-10-MgF2, Wavelength References) whose path length and pressure are 5 cm and 10 Torr, respectively. The pulses are tailored to be linearly polarized with a quarter and a half-wave plate (QWP and HWP) and pass through a wire-grid polarizer and a dichroic mirror.

For upconverting the MIR pulses to the NIR region, we use a continuous-wave 1.064-μm distributed Bragg reflector (DBR) laser (PH1064DBR200BF, Photodigm) with a linewidth of 10 MHz. The CW laser passing through a fiber-isolator is amplified with a homemade Yb-doped fiber amplifier and collimated with a collimator. The output beam goes through a free-space isolator, an HWP, and a 1-μm long-pass filter (LPF). The beam diameter and divergence are adjusted with a relay-lens pair. The beam is collinearly combined with the MIR pulses with the dichroic mirror.

### Upconversion

The combined MIR and NIR beams are focused onto a 20-mm-long PPLN waveguide (WD-3418-000-A-C-C-TEC, NTT Electronics) using a ZnSe aspheric lens with a focal length of 4.8 mm. The average power of the MIR and NIR beams measured before the ZnSe lens are a few mW and around 400 mW, respectively. The average power of the MIR pulses is intentionally decreased before coupling into the $InF_3$ fiber to suppress the undesired nonlinear optical effects in the fiber and the PPLN waveguide. Due to the DFG process in the PPLN waveguide with a polling period of 28.6 μm, a part of the 3.4-μm MIR pulses is converted to 1.5-μm NIR pulses. As shown in Supplementary Figure 1a, the center wavelength of the upconverted NIR pulses can be tuned by controlling the temperature of the PPLN crystal with a Peltier temperature controller. The NIR pulses with an average power of a few μW are collected using an aspheric lens and pass through a 1.5-μm LPF and an HWP. Then, the NIR pulses are coupled into a single-mode fiber with another aspheric lens, whose coupling efficiency is around 0.65.

### Amplification and pulse pick

The fiber-coupled NIR pulses are amplified using an Er-doped fiber amplifier with a gain up to ~30 dB (EDFA100P, Thorlabs)

and sent to a time stretcher. In large stretching cases, a pulse picker is implemented before the stretcher so that the adjacent NIR pulses do not temporally overlap each other. The pulse picker consists of a 200-MHz acousto-optic modulator (AOM) (T-M200-0.1C2J-3-F2P, Gooch&Housego) and a homemade RF driver. The RF driver generates 7-ns burst pulses with a carrier frequency of 200 MHz at a repetition rate of 10 MHz. The intensity modulation of the AOM generates NIR pulses at 10 MHz from the 80-MHz pulses. The FWHM width of the intensity modulation is 9 ns, which is determined by the width of the RF burst pulse and the rise and fall time of the AOM.

**Time-stretch**

For time-stretching, we use DCF modules (AD-SM-C-120-FC/APC-3C/3B-10, YOFC), whose dispersion parameter and insertion-loss are -0.2 ns/nm/km and -0.7 dB/km, respectively. The total length of DCF with the modules is 10-30 km, which corresponds to dispersion from -2 to -6 ns/nm. The dispersion of -12 ns/nm is achieved by the double-pass geometry of the 30-km DCF realized by implementing a fiber retro-reflector and a circulator. In large stretching cases, a Raman amplifier is additionally implemented to provide sufficient signals for detection. The Raman amplifier consists of a fiber-coupled 1.455-μm fiber-Bragg-grating (FBG)-stabilized laser (PL-FP-1455-A-A81-SA, LD-PD) with an average power of around 400 mW. The beam is coupled to the DCFs with wavelength division multiplexers (WDMs). A bidirectional pumping configuration is adopted to suppress the NIR pulses' self-phase modulation (SPM) in the DCFs while keeping the high Raman gain. The pump beam is separated by a 75:25 fiber beam splitter for bidirectional pumping. The detailed discussion about the Raman amplifier is described in Supplementary Note 3.

**Detection**

The temporally stretched NIR pulses are collimated with a collimator, spectrally filtered with 1.55-μm bandpass filters (BPF) with a bandwidth of 50 cm$^{-1}$ (12 nm), and coupled into a fiber with another collimator. The spectral filtering avoids temporal overlap of adjacent TSIR spectra and rejects amplifier noise outside the spectral bandwidth of the upconverted NIR pulses. The NIR pulses are detected with an AC-coupled 11-GHz InGaAs photodetector (RXM10AF, Thorlabs) and digitized with a high-speed 16-GHz oscilloscope (WaveMaster 816Zi-B, Teledyne LeCroy) at a sampling rate of 80 Gsamples/s.

**Retrieval of transmittance spectra**

The measurement process of a TSIR waveform is written as

$$I = \mathcal{L}\left[\left|\mathcal{F}^{-1}\left[D\sqrt{GB}\exp\left(-i\,\mathcal{K}\left[-\frac{\ln(GB)}{2}\right]\right)\right]\right|^2\right],$$

where $B \in \mathbb{R}[0,1]$ is a frequency-domain transmittance spectrum, $G \in \mathbb{R}[0,\infty)$ is a baseline spectrum, $\mathcal{K}[\cdot]$ denotes the K-K transformation, $D \in \mathbb{C}$ is the dispersion term (consisting of GDD and TOD), $\mathcal{F}^{-1}$ denotes the inverse Fourier transform, $\mathcal{L}[\cdot]$ denotes lowpass filtering, which is determined by the measured low-pass filter function (Supplementary Figure 5), and $I \in \mathbb{R}[0,\infty)$ is a measured time-domain TSIR waveform, respectively. The cut-off frequency of the low-pass filter is 11 GHz. The GDD and TOD values used in the algorithm are 2,545 ps$^2$ and -16 ps$^3$ for a 10-km DCF, 7,664 ps$^2$ and -48 ps$^3$ for a 30-km DCF, and 15,326 ps$^2$ and -96 ps$^3$ for a 60-km DCF.

We simultaneously estimate the frequency-domain transmittance spectrum $B$ and the baseline spectrum $G$ by using an iterative GD algorithm called Adam[42]. Before starting the iterations, the measured time-domain TSIR waveform

$I$ is truncated by a box-car function to restrict the frequency range for the estimation. In each iteration, we apply constraints of the sparsity and the transmittance range [0,1] to $B$ with the alternating direction method of multiplier[48] and smooth $G$ with the SG filter[41]. After the iterations, the spectral resolution of $B$ is finally adjusted to the achievable resolution (considering the width of the impulse response function) by applying the triangular apodization in the time domain. The yielded spectrum is plotted as $B$.

**Funding**

JSPS KAKENHI (20H00125, 20K05361), Precise Measurement Technology Promotion Foundation, Research Foundation for Opto-Science and Technology, Nakatani Foundation, UTEC-UTokyo FSI Research Grant Program

**Acknowledgments**

We acknowledge Yu Nagashima for the use of his equipment and Makoto Shoshin for helpful discussions.

**Author contributions**

T.I. conceived the concept of the work. K.H. designed and constructed the optical systems with the help of T.N. and T.K.. T.N. designed the optical amplifiers and the pulse picker. V.R.B. built the femtosecond mid-infrared OPO. H.S wrote the programs to control the femtosecond mid-infrared OPO and the homemade FTIR spectrometer. K.H. performed the experiments and analyzed the data. K.H. and R.H. developed the spectrum retrieval algorithm. T.I. supervised the entire work. K.H., R.H., and T.I. wrote the manuscript with inputs from the other authors.

**Disclosures**

K.H., T.N., T.K., H.S., and T.I. are the inventors of a filed patent related to UC-TSIR technique.

**Data availability**

The data provided in the manuscript are available from the corresponding author upon reasonable request.

**Supplementary Note 1: Detailed schematic of upconversion TSIR**

A detailed schematic of upconversion TSIR (UC-TSIR) is shown in Supplementary Figure 1a. Three configurations of the time stretchers with different amounts of dispersion are depicted in Supplementary Figure 1b.

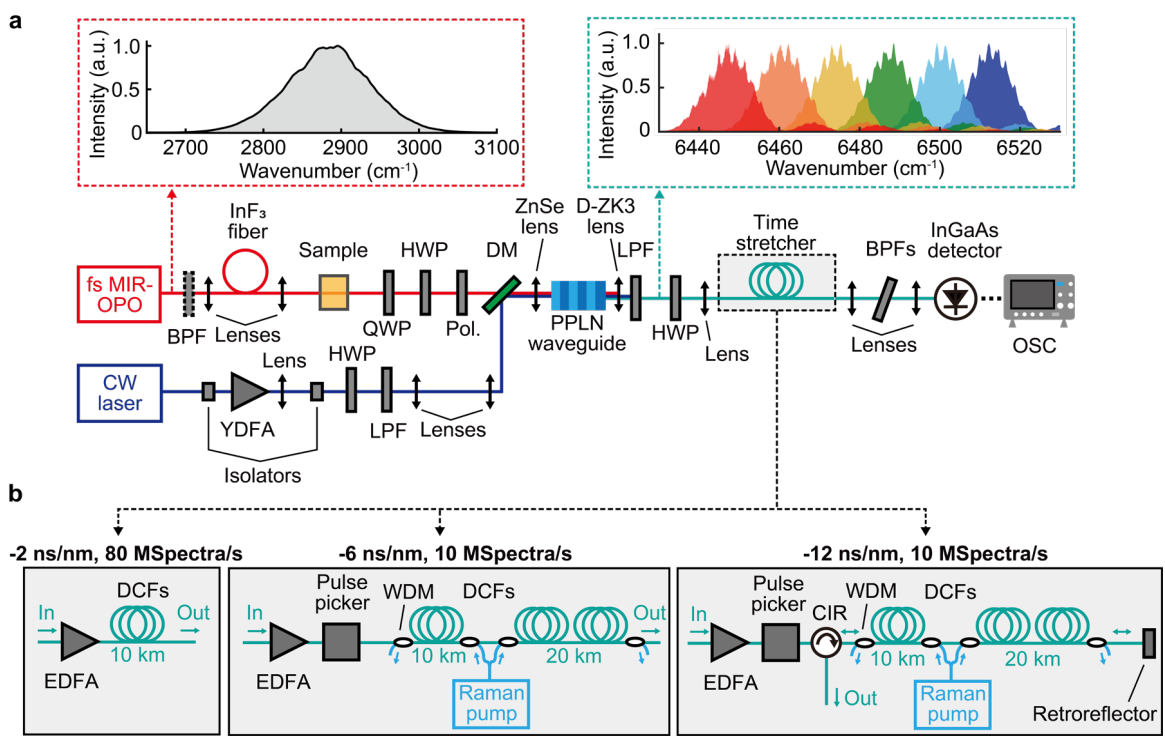

**Supplementary Figure 1: a** Detailed schematic of upconversion TSIR. The insets show a MIR spectrum generated by a fs MIR-OPO (left) and upconverted NIR spectra generated at different PPLN temperatures (13 - 63°C) (right). BPF: band-pass filter, QWP: quarter-wave plate, HWP: half-wave plate, Pol.: polarizer, DM: dichroic mirror, YDFA: Yb-doped fiber amplifier, LPF: long-pass filter, OSC: oscilloscope. **b** Time-stretcher configurations. EDFA: Er-doped fiber amplifier, DCF: dispersion-compensating fiber, WDM: wavelength division multiplexer, CIR: circulator.

**Supplementary Note 2: Nonlinear optical effects in a PPLN waveguide**

We evaluate the undesired nonlinear optical effects of MIR pulses occurring in a PPLN waveguide. Supplementary Figure 2 shows a part of upconverted $CH_4$ spectra with an average power of MIR pulses illuminated onto the PPLN waveguide of 3.6, 2.7, 1.8, 1.4, and 0.9 mW. The spectra are measured with an optical spectrum analyzer (OSA) with a spectral resolution of 0.08 cm$^{-1}$ (0.02 nm). The average power of a 1-μm CW laser for upconversion is fixed at 300 mW. As shown in the figure, the absorption lines on the upconverted spectrum distort when the average power of the MIR pulses is high. We observe that this effect does not depend on the input power of the 1-μm CW laser but on the input power and the spectral bandwidth (pulse duration) of the MIR pulses. Therefore, it is likely caused by the nonlinear optical effects (e.g., self-phase modulation (SPM)) of the MIR pulses in a PPLN waveguide. To avoid the spectral distortions, we adjust the input MIR average power to less than 1.4 mW (pulse energy of 17.5 pJ).

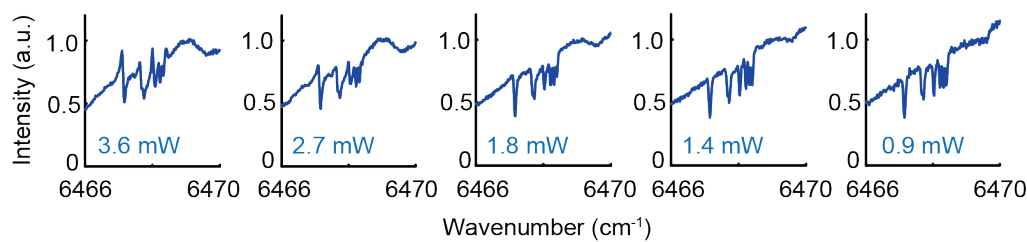

**Supplementary Figure 2:** A part of the upconverted $CH_4$ spectra measured with an OSA at a spectral resolution of 0.08 $cm^{-1}$. They are obtained at the average power of the input MIR pulses illuminated onto the PPLN waveguide of 3.6, 2.7, 1.8, 1.4, and 0.9 mW, respectively.

**Supplementary Note 3: Nonlinear optical effects in DCFs**

We also observe undesired nonlinear effects on the upconverted NIR pulses propagating in a long DCF. Supplementary Figure 3a shows a part of the upconverted $CH_4$ spectra measured after a 30-km DCF (group delay dispersion (GDD): 7,664 $ps^2$, third-order dispersion (TOD): -48 $ps^3$) that also works as a forward-pump Raman amplifier pumped with an average power of 310 or 150 mW. The repetition rate and the average power of the input NIR pulses are 10 MHz and ~20 µW (pulse energy of 2 pJ), respectively. The spectra are measured with an OSA with a spectral resolution of 0.08 $cm^{-1}$. As shown in the figure, the $CH_4$ absorption lines distort when the average pump power of the Raman amplifier is high.

To evaluate the distortion of the absorption line profile of the NIR spectra after propagating a long DCF, we simulate the SPM effects of the NIR pulses in the DCF with Raman amplification using the split-step Fourier method[1]. The calculation parameters are set to resemble the experiment shown in Supplementary Figure 3a. We assume a transform-limited 1-ps gaussian-shaped NIR pulse with a pulse energy of 2 pJ (corresponding to an average power of 20 µW at a repetition rate of 10 MHz), which interacts with molecules that have an absorption line with a linewidth of 0.01 $cm^{-1}$ (260 MHz). The pulse propagates a 30-km DCF with a forward-pump Raman amplifier (group-velocity dispersion (GVD): 255 $ps^2\,km^{-1}$, TOD per unit length: -1.6 $ps^3\,km^{-1}$, nonlinear coefficient: 5.7 $W^{-1}\,km^{-1}$, signal insertion loss: -0.7 dB $km^{-1}$, Raman-pump loss: -1.3 dB $km^{-1}$, Raman-gain coefficient: 4.7 $W^{-1}\,km^{-1}$) and is measured with a spectrometer with a spectral resolution of 0.05 $cm^{-1}$. Supplementary Figure 3b shows a part of the simulated NIR spectra after propagating the DCF with various Raman pump powers (0-400 mW), demonstrating the absorption-line distortion caused by the SPM of the NIR pulses with a high peak power gained by the Raman amplifier. Supplementary Figure 3c shows the peak power of the NIR pulses against the fiber length with the Raman pump powers of 0, 100, 200, 300, and 400 mW. These simulations show that a 10-km propagation with a peak power larger than $10^{-2}$ W causes the line-profile distortion. A backward pump configuration could alleviate the distortion.

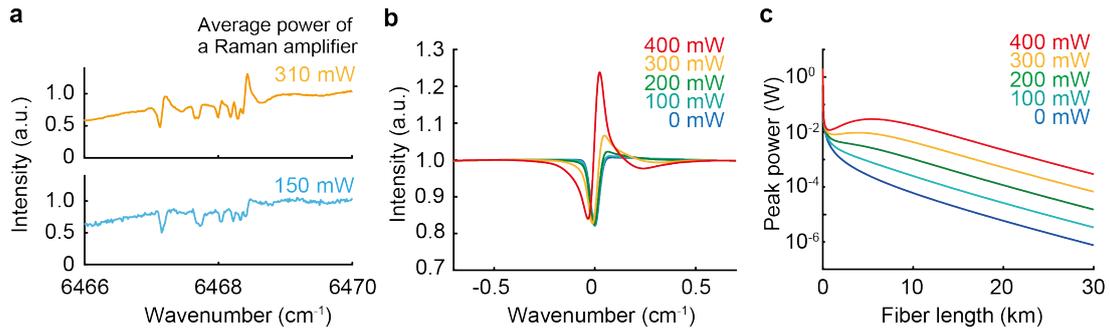

**Supplementary Figure 3:** Line-profile distortion due to the SPM of a NIR pulse in a 30-km DCF with a forward-pump Raman amplification. **a** Comparison of a part of the upconverted CH$_4$ spectra amplified with an average pump power of 310 mW (upper) and 150 mW (lower). The spectra are measured with an OSA with a spectral resolution of 0.08 cm$^{-1}$. **b** A part of the simulated spectra of a NIR pulse propagating after a 30-km DCF with average pump powers of 0, 100, 200, 300, and 400 mW. **c** Peak power of the simulated NIR pulse against the fiber length with pump powers of 0, 100, 200, 300, and 400 mW.

To avoid the distortion while keeping the high SNR, we implement a bidirectional-pump Raman amplifier, as illustrated in Supplementary Figure 1b. Supplementary Figure 4a shows a simulation result of the peak power of a NIR pulse against the fiber length in the bidirectional-pump Raman amplification, where the first part (0-10 km) is amplified with a 100-mW backward pump and the second part (10-30 km) with a 250-mW forward pump. The bidirectional-pump amplifier can keep the NIR peak power less than 10$^{-2}$ W for long-distance propagation. Supplementary Figure 4b shows a part of the spectrum with a NIR pulse propagating after the 30-km DCF with the bidirectional-pump Raman amplification, showing the distortion-less time-stretching. Finally, we experimentally demonstrate the bidirectional-pump Raman amplification. We measure upconverted CH$_4$ spectra of a NIR pulse propagating after a 30-km DCF with a bidirectional-pump Raman amplification (100-mW backward pump for the first 10 km and 250-mW forward pump for the second 20 km) using an OSA at a spectral resolution of 0.08 cm$^{-1}$ (Supplementary Figure 4c), showing no line-profile distortions. It also keeps a high average power of the NIR pulses at tens of μW (a few pJ) after the DCF.

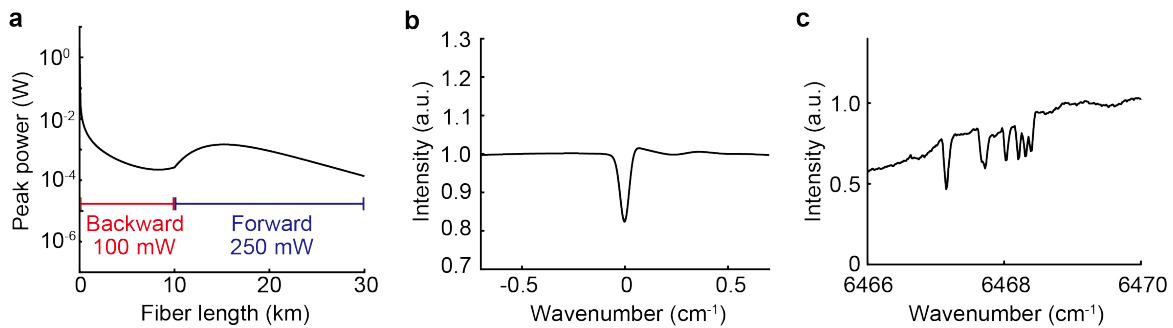

**Supplementary Figure 4:** SPM effects of a NIR pulse in a 30-km DCF with a bidirectional-pump Raman amplification. **a** Simulated peak power of a NIR pulse against the fiber length with a bidirectional Raman pump (100-mW backward pump for the first 10 km and 250-mW forward pump for the second 20 km). **b** A part of the simulated spectrum of a NIR pulse propagating after a 30-km DCF with the bidirectional-pump Raman amplification. **c** A part of the experimentally measured upconverted CH$_4$ spectrum with the bidirectional-pump Raman amplification. It is measured with an OSA with a spectral resolution of 0.08 cm$^{-1}$.

## Supplementary Note 4: Characterization of the InGaAs photodetector

We evaluate the frequency response of the InGaAs photodetector used in our experiment. Supplementary Figure 5a shows the impulse response obtained by measuring an unstretched NIR pulse with the detector, showing the FWHM width of 49 ps. Supplementary Figure 5b shows the frequency response of the detector obtained by Fourier-transforming the train of the temporal impulse responses. The -3dB bandwidth of the detector is evaluated as 11 GHz.

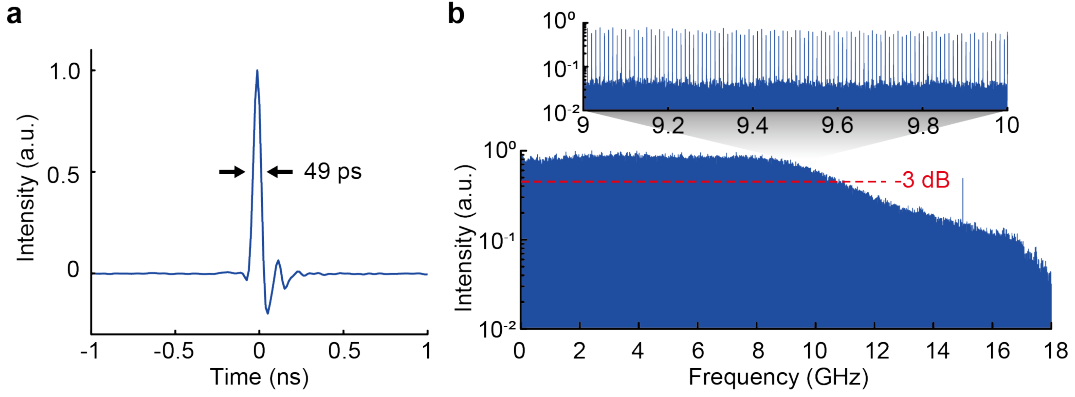

**Supplementary Figure 5:** Characterization of the InGaAs photodetector used in our experiment. **a** An impulse response of the detector. The FWHM width is 49 ps. **b** Frequency response of the detector. The -3dB bandwidth is 11 GHz.

## Supplementary Note 5: Theoretical description of a TSIR spectrum

We theoretically describe a TSIR waveform to show why the absorption lines on the spectrum become dispersive. In this description, we only consider GDD as a dispersion for time-stretching for simplicity. We add GDD to an input electric field, $E_{\text{in}}(\omega)$, in the frequency domain as

$$E_{\text{out}}(\omega) = E_{\text{in}}(\omega)\exp\left\{i\frac{\beta_2 L}{2}(\omega - \omega_0)^2\right\}, \tag{S1}$$

where $\omega$, $\omega_0$, $\beta_2$, and $L$ denote an angular frequency, a center angular frequency, GVD, and a fiber length, respectively. $E_{\text{in}}(\omega)$ and $\exp\left\{i\frac{\beta_2 L}{2}(\omega - \omega_0)^2\right\}$ correspond to $\sqrt{GB}\exp\left(-i\mathcal{K}\left[-\frac{\ln(GB)}{2}\right]\right)$ and $D$ in Methods in the main text, respectively. The time-domain electric field, $e_{\text{out}}(t)$, is written as

$$\begin{aligned} e_{\text{out}}(t) &= \mathcal{F}^{-1}[E_{\text{out}}(\omega)] \\ &= \mathcal{F}^{-1}\left[E_{\text{in}}(\omega)\exp\left\{i\frac{\beta_2 L}{2}(\omega - \omega_0)^2\right\}\right] \\ &= \mathcal{F}^{-1}[E_{\text{in}}(\omega)]\otimes\mathcal{F}^{-1}\left[\exp\left\{i\frac{\beta_2 L}{2}(\omega - \omega_0)^2\right\}\right], \end{aligned} \tag{S2}$$

where $t$, $\mathcal{F}^{-1}$, and $\otimes$ denote time, the symbols of operations of inverse Fourier transform and convolution,

respectively. $\mathcal{F}^{-1}[E_{in}(\omega)]$ and $\mathcal{F}^{-1}\left[\exp\left\{i\frac{\beta_2 L}{2}(\omega-\omega_0)^2\right\}\right]$ are individually described as

$$\mathcal{F}^{-1}\{E_{in}(\omega)\} = \frac{1}{2\pi}\int \{E_{in}(\omega)\}\exp(i\omega t)d\omega$$
$$= e_{in}(t), \quad (S3)$$

and

$$\mathcal{F}^{-1}\left[\exp\left\{i\frac{\beta_2 L}{2}(\omega-\omega_0)^2\right\}\right] = \frac{1}{2\pi}\int \left[\exp\left\{i\frac{\beta_2 L}{2}(\omega-\omega_0)^2\right\}\right]\exp(i\omega t)d\omega$$
$$= \frac{1}{\sqrt{2\pi\beta_2 L}}\exp\left\{i\left(\frac{\beta_2 L}{2}\omega_0^2 + \frac{\pi}{4}\right)\right\}\exp\left\{-i\frac{(t-\beta_2 L\omega_0)^2}{2\beta_2 L}\right\}, \quad (S4)$$

respectively, where $e_{in}(t)$ denotes the time-domain input electric field. Then, $e_{out}(t)$ is described as

$$e_{out}(t) = e_{in}(t)\otimes A\exp\left\{-i\frac{(t-\beta_2 L\omega_0)^2}{2\beta_2 L}\right\}$$
$$= A\int e_{in}(t')\exp\left\{-i\frac{(t-\beta_2 L\omega_0-t')^2}{2\beta_2 L}\right\}dt', \quad (S5)$$

where $t'$ denotes time, and $A$ is defined as $A = \frac{1}{\sqrt{2\pi\beta_2 L}}\exp\left\{i\left(\frac{\beta_2 L}{2}\omega_0^2 + \frac{\pi}{4}\right)\right\}$. Using the variable, $\tau = t - \beta_2 L\omega_0$, Equation (S5) is rewritten as

$$e_{out}(\tau) = A\int e_{in}(t')\exp\left\{-i\frac{(\tau-t')^2}{2\beta_2 L}\right\}dt'. \quad (S6)$$

Therefore, a TSIR waveform, $I(\tau)$, is expressed as

$$I(\tau) = |e_{out}(\tau)|^2$$
$$= |A|^2\left|\int e_{in}(t')\exp\left\{-i\frac{(\tau-t')^2}{2\beta_2 L}\right\}dt'\right|^2 \quad (S7)$$
$$= |A|^2\left|\int e_{in}(t')\exp\left(-i\frac{t'^2}{2\beta_2 L}\right)\exp\left(i\frac{\tau}{\beta_2 L}t'\right)dt'\right|^2.$$

When $\left|\frac{t'^2}{2\beta_2 L}\right| \ll 1$, $I(\tau)$ is approximately written as

$$I(\tau) \approx |A|^2\left|\int e_{in}(t')\exp\left\{-i\left(-\frac{\tau}{\beta_2 L}\right)t'\right\}dt'\right|^2 \quad (S8)$$

$$= |A|^2 \left| \int e_{in}(t') \exp(-i\omega' t') dt' \right|^2$$

$$\propto |\mathcal{F}\{e_{in}(t')\}|^2$$

$$= |E_{in}(\omega')|^2,$$

where $\mathcal{F}$ denotes the symbol of Fourier transform operation, and $\omega'$ is defined as $\omega' = -\frac{\tau}{\beta_2 L}$. In this regime, a TSIR waveform in the time domain represents the spectrum in the frequency domain (far-field regime). If the approximation is invalid, $e_{in}(t')$ is modulated by the term, $\exp\left(-i\frac{t'^2}{2\beta_2 L}\right)$, which yields the dispersive shape on the absorption lines (near-field regime).

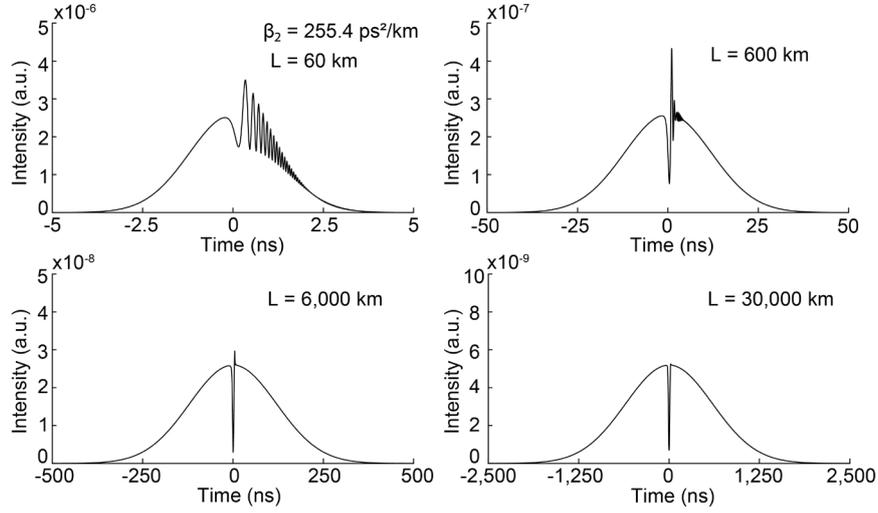

**Supplementary Figure 6:** TSIR waveforms calculated with various amounts of dispersion. We calculate a 1-cm$^{-1}$-width gaussian-shape spectrum with a 0.01-cm$^{-1}$-width absorption line and temporally stretch it using fibers with a GVD of 255.4 ps$^2$/km at four different lengths (60, 600, 6000, and 30000 km). $\beta_2$: GVD, $L$: fiber length.

We simulate TSIR waveforms with an absorption line using the above-described equations. Supplementary Figure 6 shows TSIR waveforms calculated by adding various amounts of dispersions to a 1-cm$^{-1}$-width gaussian-shaped spectrum with a 0.01-cm$^{-1}$-width absorption line. As a dispersive medium, we assume a DCF with a GVD of 255.4 ps$^2$/km at four different lengths (60, 600, 6,000, and 30,000 km). We find a 10$^4$-km DCF is necessary for a TSIR waveform to reach the far-field regime, which is not experimentally realistic to be used. Therefore, it is practical to measure a TSIR waveform in the near-field regime with a moderate-length fiber and make a computational spectral correction. We note that the fiber length of 1 km is enough to reach the far-field regime when measuring condensed-phase infrared spectra with a physical linewidth of ~3 cm$^{-1}$.

**Supplementary Note 6: Wavenumber instability of TSIR spectroscopy**

We evaluate the wavenumber instability of TSIR spectroscopy by measuring the peak-position fluctuation of continuously measured single-shot TSIR spectra. The lower panel in Supplementary Figure 7 shows the temporal variation of the TSIR spectra (0-9 μs) sequentially measured at 10 Mspectra/s with the 30-km DCF (total dispersion of -6 ns/nm). Each TSIR waveform is segmented, and the temporal position is calibrated with the peak at 43.11 ns (shown in the upper panel of Supplementary Figure 7). The standard deviation of the peak positions around 56.16 ns is 9 ps, which is shorter than the sampling period of the oscilloscope (12.5 ps).

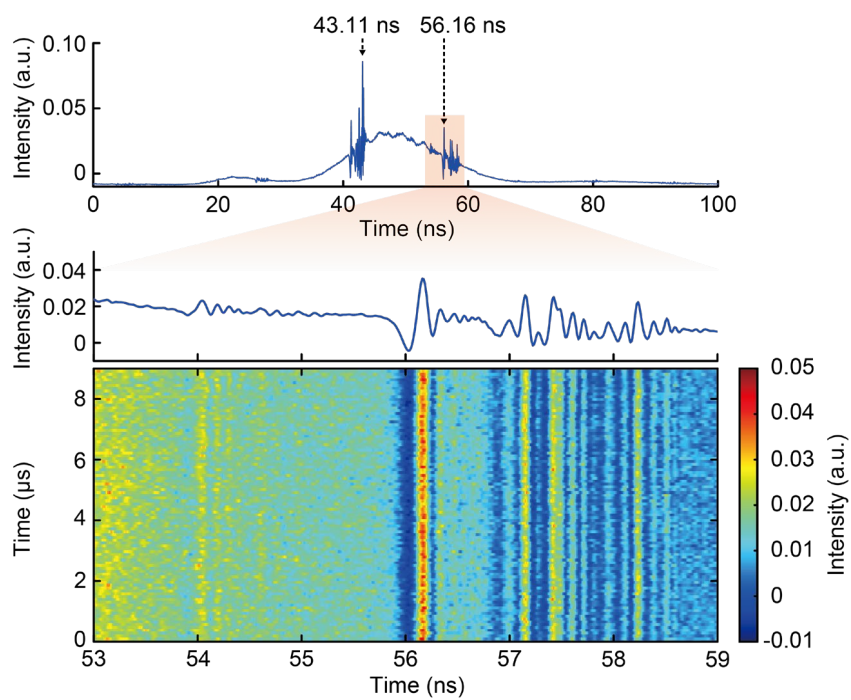

**Supplementary Figure 7:** Wavenumber instability of TSIR spectroscopy. The upper and middle panels show the averaged TSIR spectrum measured at 10 Mspectra/s with the 30-km DCF and the enlarged view of the spectrum in the time frame of 53-59 ns, respectively. The lower panel represents the temporal variation of single-shot TSIR spectra in the time frame of 53-59 ns sequentially measured at 10 Mspectra/s.

**Supplementary Note 7: Relative wavenumber accuracy of TSIR spectroscopy**

We evaluate how the third-order dispersion (TOD) affects the relative wavenumber accuracy of a retrieved spectrum. Supplementary Figure 8a shows transmittance spectra of $CH_4$ molecules retrieved from 180-times averaged TSIR waveforms stretched with a dispersion of -12 ns/nm. The red plot represents the case where both GDD (=15,326 $ps^2$) and TOD (= -96 $ps^3$) values are taken into account for the GD algorithm, while the orange plot represents the case where only the GDD value is taken into account. The x-axis is the MIR-wavenumber downconverted from the NIR-wavenumber. As shown in the inset of the figure, the peak position of the orange plot is shifted by 0.027 $cm^{-1}$ at 2,916 $cm^{-1}$. Supplementary Figure 8b shows the difference in peak position ($d_{wav}$) against the peak positions of the calculated spectrum from the HITRAN database. The red dots represent the case where both GDD and TOD values are taken into account for the algorithm, while the orange dots represent the case where only the GDD value is taken into account. The absolute maxima of $d_{\text{wav}}$ are 0.007 and 0.026 cm$^{-1}$ in the former and the latter cases, respectively. It verifies taking into account the TOD is essential to improve the relative wavenumber accuracy.

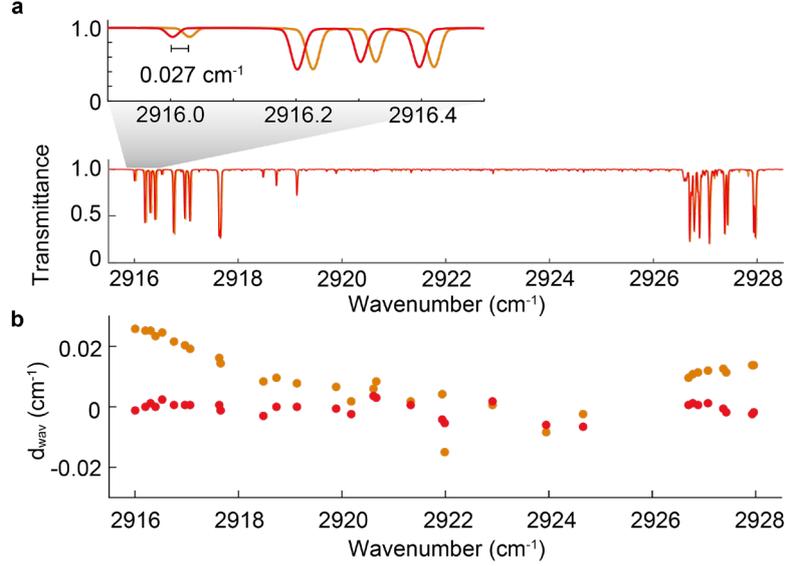

**Supplementary Figure 8:** Relative wavenumber accuracy of TSIR spectroscopy. **a** Transmittance spectra of CH$_4$ molecules retrieved from TSIR waveforms stretched with a dispersion of -12 ns/nm. The red and orange plots represent the cases where the GD algorithm includes GDD and TOD values and a GDD value only, respectively. The inset shows a zoom-in view of the spectra. **b** Difference in absorption-peak position between the retrieved spectra and the calculated spectrum from the HITRAN database ($d_{\text{wav}}$). The red and orange dots represent the cases where the GD algorithm includes GDD and TOD values and a GDD value only, respectively.

**Supplementary Note 8: SNR comparison between TSIR and FTIR spectra**
We theoretically evaluate the SNR of a TSIR spectrum and compare it with that of an FTIR spectrum. We individually derive the SNR of TSIR and FTIR spectra and then provide the ratio between them.

First, we derive the SNR of a TSIR spectrum. The SNR of a temporal intensity waveform, $I(t)$, is defined as

$$SNR_{\text{t}}(t) = \frac{I(t)}{\epsilon_{\text{t}}}, \tag{S9}$$

where $t$ and $\epsilon_{\text{t}}$ denote time and noise (r.m.s. error)$^2$, respectively. For simplicity, we assume the white noise and a flat-top TSIR spectrum (constant signal intensity with respect to time) with a duty cycle of 1. The SNR of a TSIR spectrum, $SNR_{\text{TS}}$, can be written as

$$SNR_{\text{TS}} = \frac{P}{\sigma(P)\sqrt{f_{\text{BW}}}}, \tag{S10}$$

where $P$, $\sigma(P)$, and $f_{\text{BW}}$ denote an average detection power, a system's overall noise-equivalent power (NEP), and an RF bandwidth of the system, respectively. The overall NEP depends on the average detection power because it contains various noises (e.g., detector noise, shot noise, or amplified spontaneous emission (ASE) noise). The RF bandwidth is mainly determined by the detector and the digitizer. Throughout this comparison, we set the sampling frequency is double the RF bandwidth, $f_{BW}$. In this condition, the SNR can be written as

$$SNR_{TS} = \sqrt{\frac{2T}{N}} \frac{P}{\sigma(P)}, \qquad (S11)$$

where $T$ and $N$ denote a measurement time and the number of the sampling data points, respectively. Since the number of spectral elements, $M$, of TSIR measurement is equal to the number of the sampling data points, $N$, the SNR can be written as

$$SNR_{TS} = \sqrt{\frac{2T}{M}} \frac{P}{\sigma(P)}. \qquad (S12)$$

The equation shows the SNR of a TSIR spectrum is proportional to the square root of the measurement time (spectral acquisition time) and inversely proportional to the square root of the number of spectral elements. If there is a duty cycle, $D_{TS} = \frac{M}{N}$ ($0 \leq D_{TS} \leq 1$), it may be written as

$$SNR_{TS} = \sqrt{\frac{2T}{M}} \frac{1}{\sqrt{D_{TS}}} \frac{P}{\sigma\left(\frac{P}{D_{TS}}\right)}. \qquad (S13)$$

Next, we derive the SNR of an FTIR spectrum. We define the SNR of a spectral density, $B(f)$, as

$$SNR_f(f) = \frac{B(f)}{\epsilon_f}, \qquad (S14)$$

where $f$ and $\epsilon_f$ denote frequency and noise, respectively. We first describe the SNR of a double-sided interferogram, $I(t)$, by the spectral density by following the procedure given in the literature[2]. The center burst of an interferogram, $I(0)$, can be written as

$$I(0) = \sum_{j=1}^{N} B(f_j) \, \delta f \qquad (S15)$$
$$= \bar{B} \cdot N \cdot \delta f,$$

where $\bar{B}$ and $\delta f$ denote the average spectral density and the spectral interval between adjacent spectral elements, respectively. Considering the energy conservation of Fourier transformation[2], the relation between the noises in the time and frequency domains is written as

$$(\epsilon_t)^2 \cdot \delta t \cdot N = \left(\epsilon_{f,tot}\right)^2 \cdot \delta f \cdot N, \qquad (S16)$$

where $\delta t$ and $\epsilon_{f,tot}$ denote the sampling period in the time domain and the total noise in the frequency domain. Since only the real part of the Fourier transformation contributes to the FTIR noise ($\epsilon_f = \epsilon_{f,tot}/\sqrt{2}$), $\epsilon_t$ can be

written as

$$\epsilon_t = \sqrt{2}\epsilon_f \sqrt{\frac{\delta f}{\delta t}}. \tag{S17}$$

Therefore, the SNR of the interferogram's center burst, $SNR_t(0)$, is written as

$$\begin{aligned} SNR_t(0) &= \frac{I(0)}{\epsilon_t} \\ &= \frac{\bar{B} \cdot N \cdot \delta f}{\sqrt{2}\epsilon_f \sqrt{\frac{\delta f}{\delta t}}} \\ &= \sqrt{\frac{N}{2}} \frac{\bar{B}}{\epsilon_f}. \end{aligned} \tag{S18}$$

Here, we use the relation $\delta t \delta f = N^{-1}$. We can describe the SNR of an FTIR spectrum as

$$SNR_{FT}(f) = \sqrt{\frac{2}{N}} \frac{B(f)}{\bar{B}} SNR_t(0). \tag{S19}$$

We calculate the SNR of the interferogram's center burst, $SNR_t(0)$. We assume that an interferogram with the visibility of 1 is AC-coupled and measured at a measurement time of $T$ with the number of sampling data points of $N$. Since the two beams from the sample and reference arms constructively interfere at $t = 0$, $SNR_t(0)$ can be written as

$$\begin{aligned} SNR_t(0) &= \frac{P}{\sigma(P)\sqrt{f_{BW}}} \\ &= \sqrt{\frac{2T}{N}} \frac{P}{\sigma(P)}. \end{aligned} \tag{S20}$$

Considering the Nyquist-limited spectral measurement ($N/2 = M$) of a flat-top spectrum ($\bar{B} = B(f)$), the SNR of FTIR is described as

$$\begin{aligned} SNR_{FT} &= \frac{1}{\sqrt{M}} SNR_t(0) \\ &= \frac{\sqrt{T}}{M} \frac{P}{\sigma(P)}. \end{aligned} \tag{S21}$$

Now, we compare the SNR of TSIR (S12) and FTIR (S21) spectra. The SNR ratio of a TSIR spectrum and an FTIR spectrum, $R_{TS/FT} = \frac{SNR_{TS}}{SNR_{FT}}$, is expressed as

$$R_{\text{TS/FT}} = \sqrt{2M}. \tag{S22}$$

It shows a TSIR spectrum has $\sqrt{2M}$-times higher SNR than an FTIR spectrum under the same measurement time. In other words, TSIR can take spectra with the same SNR at a $2M$-times higher spectral measurement rate than FTIR.

Finally, we compare the SNR of TSIR and FTIR, assuming that we can fully utilize the detector's dynamic range. Since half of the average detection power is used for the DC components of an FTIR interferogram, the dynamic-range-limited average detection power of FTIR is half of TSIR ($P_{\text{TS}} = 2P_{\text{FT}}$). To compare the SNR values under different average detection powers, we describe the overall NEP for each noise component as

$$\sigma_{\text{TS(FT)}}(P_{\text{TS(FT)}}) = \sqrt{NEP_{\text{det}}^2 + \left(\sqrt{\frac{2P_{\text{TS(FT)}}h\nu}{\eta}}\right)^2 + \left(\sqrt{(Amp)P_{\text{TS(FT)}}^2}\right)^2}, \tag{S23}$$

where $P_{\text{TS(FT)}}$, $NEP_{\text{det}}$, $h$, $\nu$, $\eta$, and $Amp$ denote average detection power for TSIR (FTIR), detector NEP, the Planck constant, optical frequency, the quantum efficiency of a detector, and amplifier noise of an amplifier, respectively. The three terms, $NEP_{\text{det}}$, $\sqrt{\frac{2P_{\text{TS(FT)}}h\nu}{\eta}}$, and $\sqrt{(Amp)P_{\text{TS(FT)}}^2}$, correspond to the NEP dominated by the detector noise, the shot noise, and the amplifier noise, respectively. We assume the digitizer's dynamic range is sufficiently high and does not limit the SNR. With the relation $P_{\text{TS}} = 2P_{\text{FT}} = 2P$, the SNR ratio is expressed as

$$\begin{aligned} R_{\text{TS/FT}} &= \frac{P_{\text{TS}}}{P_{\text{FT}}} \frac{\sigma_{\text{FT}}(P_{\text{FT}})}{\sigma_{\text{TS}}(P_{\text{TS}})} \sqrt{2M} \\ &= \left\{ 2 \frac{\sigma_{\text{FT}}(P)}{\sigma_{\text{TS}}(2P)} \right\} \sqrt{2M} \\ &= \alpha\sqrt{2M}. \end{aligned} \tag{S24}$$

The constant value $\alpha$ varies from 1 to 2 depending on the noise condition. For example, if the noise is dominated by the detector noise, the shot noise, or the amplifier noise, it becomes 2, $\sqrt{2}$, or 1, respectively.

**Supplementary Note 9: SNR evaluation of upconversion TSIR spectroscopy**
We evaluate the SNR of upconversion TSIR spectroscopy using transmittance TSIR waveforms measured at a rate of 10 MHz. The black dots in Supplementary Figure 9 show the SNR of a single-shot TSIR waveform (measured with 30-km DCF) against the average power detected with an InGaAs detector. The average detection power is changed right before the photodetector. The SNR is evaluated from the standard deviation of a normalized waveform where no large peaks exist. It linearly increases against the average detection power at a lower power level while being saturated around ten above ~5 µW.

To compare them with theory, we also plot the calculated SNR. The SNR lines are calculated assuming a flat-top spectrum at a spectral measurement rate of 10 MHz, a duty cycle of 0.16 at FWHM, and an RF bandwidth of 11 GHz.

We also assume the detector NEP of $15.5\ pW\ Hz^{-\frac{1}{2}}$, the wavenumber of 6,471 cm$^{-1}$, and the detector quantum efficiency of 0.72. The blue line in the figure represents a detector-noise-limited SNR calculated from the NEP of the InGaAs photodetector and the digitizer, which linearly increases against the average detection power. It agrees well with the measured SNR at a lower average detection power than 5 μW. The green line represents the shot-noise-limited SNR of the current system, which is determined by the number of photons (corresponding to an average power of 250 nW at 10 MHz) before the optical amplification with the EDFA. The measured SNR above 5 μW is slightly lower than the shot-noise-limited SNR. We note that the shot-noise-limited SNR described here, SNR$_{shot-UC}$, does not depend on the average detection power. We attribute the additional noise to the optical amplifier, which also gives a constant SNR against the average detection power with the photodetector. The orange line shows the upper limit of an input power where the detector linearly responds without saturation.

The SNR can be improved by making some modifications to the system. One can realize the detector-noise limited SNR up to ~90 at a higher average detection power above 5 μW by increasing the number of upconverted photons before the optical amplification, which improves the shot-noise-limited SNR and can also reduce the optical amplification noise. We expect the number of upconverted photons can increase by injecting a higher power of the 1-μm laser into the PPLN waveguide. There is a potential to increase the SNR further by using a lower-noise detector. The red line in Supplementary Figure 9 represents the shot-noise-limited SNR determined by the number of photons captured by a photodetector, SNR$_{shot}$, which linearly increases against the square root of the average detection power.

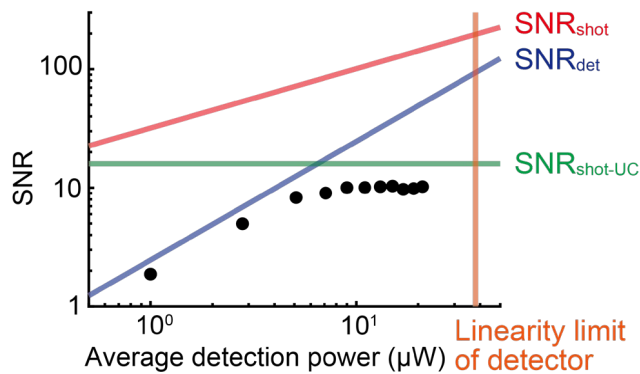

**Supplementary Figure 9:** Single-shot SNR of a TSIR waveform against an average power captured by an InGaAs photodetector. The black dots represent the SNR of the single-shot normalized TSIR waveforms measured at 10 MSpectra/s with 30-km DCF. The blue, green, and red lines represent the calculated SNR dominated by the detector noise (SNR$_{det}$), the shot noise determined by the number of photons before the optical amplification with an EDFA (SNR$_{shot-UC}$), and the shot noise determined by the number of photons captured by a photodetector (SNR$_{shot}$), respectively. The orange line represents the upper limit of the average detection power determined by the detector's linearity. The SNR linearly increases against the average detection power due to the detector noise, while it is saturated around ten due to the shot noise determined by the number of photons before the EDFA and the amplifier noise.

**Supplementary References**